\documentstyle{article}

\def\be{\begin{equation}}
\def\ee{\end{equation}}
\def\bea{\begin{eqnarray}}
\def\eea{\end{eqnarray}}

\newcommand{\bc}{\begin{center}}
\newcommand{\ec}{\end{center}}

\newcommand{\barr}{\begin{array}}
\newcommand{\bey}{\begin{eqnarray}}
\newcommand{\ear}{\end{array}}
\newcommand{\eey}{\end{eqnarray}}

\newcommand{\ssty}{\scriptstyle}
\newcommand{\sssty}{\scriptscriptstyle}

\newsavebox{\ipiu}
\newsavebox{\imen}

\sbox{\ipiu}{$\ssty i \sssty +1$}
\sbox{\imen}{$\ssty i \sssty -1$}

\begin{document}

\vspace{2cm}

\begin{center}
{\bf From Heisenberg to Einstein? Recollections and afterthoughts on the birth of string theory\footnote{Contribution to the collective volume ``The birth of String Theory'', edited by A. Cappelli, E. Castellani, F. Colomo and P. Di Vecchia.}}
\end{center}

\vspace{1cm}

\begin{center}
 Renato  MUSTO \\
Dipartimento di Scienze Fisiche,
Universit\`a di Napoli Federico II,\\
and INFN, Sez. di Napoli,\\
80126, Napoli, ITALY\\
e-mail: musto@na.infn.it
\end{center}


\begin{center}
\bf Abstract
\end{center}

A few recollections and afterthoughts on the development of the string picture of fundamental interactions out of the S-matrix program.

\vspace{2cm}

1968. We were young. Revolution was in the air. Gabriele Veneziano's paper \cite{Veneziano} was an event in that eventful year. It was immediatly clear that it was an important result, but very few people would have thought of it as the beginning of an enterely new theoretical approach that was going to grow in the shell of the old one.  Since then, I have been many times in and out of active research in physics, engaged, 
in the meantime,  in different problems in the history of ideas. It seems, then,  appropriate to me to try to combine in this paper my personal recollections of the days of the birth of string theory with a few observations on a change  in perspective that occurred in theoretical physics. It is  a change  that we expected in society and  that we have witnessed in the course of the years in physics.

It is enough to look at the high energy physics titles of the W. A. Benjamin Inc.   series {\it Frontiers in Physics}, that started in 1961, to get the feeling of the early Sixties: in the first  year Goffrey F. Chew's {\it  S Matrix Theory  of Strong Interactions};  in 1963 {\it Regge poles and  S Matrix} by S. C. Frautschi, {\it Mandalstam Theory and Regge Poles}
by R. Omn\`{e}s and M. Froissart together with {\it Complex Angular Momenta and Particle Physics } by E.J. Squires; in 1964 {\it Strong-interaction Physics} by M. Jacob and G. F. Chew\footnote{    Chew's {\it Analytic  S Matrix} with its 
openly programmatic subtitle {\it A Basis for Nuclear Democracy} was published in 1966 again by Benjamin.}. In order to complete the list one has to add to these  S-matrix inspired titles only few books, among them {\it The Eightfold Way} by M. Gell-Mann and Y. Ne'eman.

There were  different and connected reasons for the dominant role of S-matrix and for the increasing suspicion of field theory methods. 
The mathematical foundation of field theory appeared unsound. For many physicists field theory 
meant only Feynman graphs. The renormalization program, even if successful in  curing   the problem of infinities present in perturbative expressions,    appeared  a mathematical trick, as its physical meaning was not yet understood. As  is clearly expressed in a classic textbook:{ \it ``There is as yet no logically consistent and complete relativistic quantum field theory. We shall see that the existing theory introduces new physical features into the nature of the description of particle states, which acquire some of the features of field theory. The theory is, however, largely constructed on the pattern of ordinary quantum mechanics and makes use of the latter's concepts. This structure of the theory has yielded good results in quantum electrodynamics. The lack of complete logical consistency in the theory is shown by the occurrence of divergent expressions when the mathematical  formalism is directly applied,  although there are quite well-defined ways of eliminating these divergences. Neverthless, such methods remain, to a considerable extent, semipratical rules, and our confidence on the correctness of the results is ultimately based only on their excellent agreement with experiment, not on the internal consistency of logical  ordering of the fundamental principles of the theory."} \cite{Landau}. Even worse, the  presence of Landau pole, found in the behaviour of what today is called the running of the coupling constant in quantum electrodynamics,  appeared to be a fatal blow to the full consistency of the theory. To quote again from Landau and Lifsitz: ``{\it Thus we reach the important conclusion that quantum electrodynamics  as a theory with weak interaction is essentially incomplete. Yet the whole formalism of the existing  theory depends on the possibility of regarding the electromagnetic interaction as a small perturbation}" \cite{Landau1}. To fully understand what that meant, one must realize that the results of quantum electrodynamics were thought of at that time as having  universal validity in quantum field theory\footnote{An interesting discussion on these points may be found in ref. \cite{Gross}.}.

But it was in the realm of strong interactions, where the  most intense experimental activity was performed, that the use of quantum field theory appeared more problematic. At first, the scheme of quantum electrodynamics appeared successful with Yukawa description of nuclear forces in terms of nucleon-pion interaction. But, then, the increasing number of resonances that were found, made such a scheme useless. Furthermore the strength of the coupling was a fundamental obstacle, as non-perturbative methods were not available. To cope with these problems, a new philosophical perspective was developed. Instead of starting from a fundamental set of fields, the attempt was to construct a self-consistent {\it bootstrap} scheme ``{\it with all strongly interacting particles (hadrons) enjoying equivalent status in a `nuclear democracy'}" \cite{Chew66a}. The correct framework for this program  was not quantum field theory, which requires the choice of a fundamental set of fields and the construction of a hamiltonian space-time evolution, but the  S-matrix approach of Wheeler and Heisenberg: ``{\it The simple framework  of S-matrix theory and the restricted set of questions that it presumes to answer constitutes a major advantage over quantum field theory. The latter is burdened with a superstructure, inherited from classical electromagnetic theory, that seems designed to answer a host of experimentally unanswerable questions.}" \cite{Chew66a}\footnote{The term `S-Matrix'  has been introduced by Wheeler in 1937 \cite{Wheeler},  the S-matrix program has been developped by Heisenberg starting in 1943 \cite{Heis1943}.  For a through discussion of the  developments of the  S-matrix philosophy see ref. \cite{Cushing}.}. 

By the time of the revival of his  S-matrix  program  Heisenberg had long changed his point of view, looking for a fundamental {\it nonlinear } field theory:
``{\it It is perhaps not exaggerated to say that the study of the  S-matrix is a very  useful method for deriving relevant results for collision processes by going around fundamental problems. But these problems should be solved... The  S-matrix is an important but very complicated mathematical quantity that should be derived from the fundamental field equations...}" \cite{Heis1957}. According to Chew, this change in perspective was due to the weakness of the original formulation of the  S-matrix theory that lacked dynamical content: "{\it Heisenberg lost interest probably because in the forties he lacked the full analytic continuation that is required to give the  S-matrix dynamical content}" \cite{Chew61}.

In the second half of the Fifties analytic properties of scattering amplitudes  were extensively studied in the framework of field theory and at the beginning  of the Sixties Regge poles and analytic continuation   in angular momentum were taken  from  potential scattering into relatavistic high energy physics. These were the two ingredients that, according to Chew, were needed to make S-matrix a full dynamical theory. It is worth noticing that in 1961, at the Twelfth Solvay Conference, on answering to a comment by Chew,  Heisenberg recalled his  reason for giving up the construction of a pure S-Matrix theory: ``{\it ... when one constructs a unitary S-Matrix from simple assumptions (like  a hermitian $\eta$ matrix by assuming $S=e^{i\eta}$), these S-matrices always become non analytical at places where they ought be analytical. But I found it very difficult to construct analytic S-matrices}." \cite{Solvay}. But it was exactly the difficulty encountered in constructing an S-matrix satisfying the requirements of unitarity and analyticity and exibiting the expected Regge behaviour that was the reason for hoping  in an extraordinary predictive power of such a program, once accomplished.

In the spirit of Heisenberg's ideas of the Fourties the bootstrap program was giving up the possibility of a space-time description of microscopic phenomena in favour of an even more ambitious goal, namely to obtain complete knowledge of masses and coupling constants, perhaps in terms of a unique constant with the dimension of a length:  "{\it The belief is growing that  a successful  theory of strong interactions will not tolerate arbitrary parameters; perhaps no arbitrariness of any kind is possible...}"\cite{Chew66a}. This extremely  attractive, though certainly disputable,  ideal of total predictability, so much in tune with the spirit of the time,  is, probably, the strongest heritage that string theory has received from  the S-matrix program.   
  
It was impossible for a graduate student to escape the fascination of these ideas in a period  when it was claimed that ``the times they are a-changin'~" and that ``the old road is rapidly agin'~" \cite{Dylan}. Youth appeared as a value by itself and quantum field theory appeared old, at least as it was taught to me. (Even worse, General Relativity  was taught as a branch of mathematics.) I was  fascinated by  the idea of giving up a space-time description that seemed to be  a fake theatrical perspective, deprived of any dynamical  content. I was attracted by the possibility of being in touch with the latest  developments in high-energy physics, but I felt quite uneasy with the lack of a strong theoretical structure and the continuous guesswork that  S-matrix approach required. I was much more confident in the algebraic methods that, with  the success of the Eightfold Way,  became a promising substitute for a lacking sound dynamical principle. Symmetry and non-symmetry group had proven  very effective in potential theory, as in the case of the Hydrogen atom \cite{hydrogen}; the algebra of currents, that Murray  Gell-Mann had abstracted from field theory \cite{Gell-Mann}, was opening a new path for the understanding of strong interactions. Under the guidance of Lochlain O' Rafeartaaigh I derived  algebraic relations  in the bootstrap approach, mainly by using  finite-energy sum rules.  

Then everything changed. Dolen, Horn and Schmid duality \cite{Dolen} suggested that resonance and Regge pole contributions shouldn't be added and Gabriele Veneziano implemented explicitly the idea in the case of a  four-point scalar amplitude \cite{Veneziano}. He showed that the right  starting point of the S-matrix program was {\it not}  a unitary S-matrix but a crossing-symmetric  amplitude made up by an infinite sum of zero-width resonances lying on linearly rising Regge trajectories building up Regge behaviour at high energy. While the attempt of saturating finite-energy sum rules to get numerical or algebraic relations involved only a finite number of  states, the essential feature of Veneziano amplitude  was the presence of an infinite number of states. The theory of the dual resonance model (DRM) had started.  

By June 1969, when Sergio Fubini visited Napoli, much progress had  already been made in DRM\footnote{For a discussion of results and methods of the first phase of DRM and string theory see, for instance,  ref. \cite{Di Vecchia}.}. And Sergio  shared with us not only his knowledge of the latest developments but also his enthusiasm and his vision of the new theorethical structure that was beeing unraveled. On  that occasion I met some of the brillant young physicists with whom I became coworker and friend. It was then that I started my activity in DRM.

In 1971 I joined Emilio Del Giudice, Paolo Di Vecchia, Sergio Fubini at MIT, where they had completed their important work on transverse physical states operators \cite{Del Giudice}.  The structure of DRM had been largely clarified but, in the meantime, the physical picture of strong interactions had been drastically changed by the deep inelastic scattering experiments, probing the short distance structure of nucleons. The question of the day was to explain the almost free behaviour of elementary constituents inside the nucleon. Could DRM help in understanding this new physics? In order to clarify this point, we classified multi-parton states by means of a subgroup of the conformal group, obtaining  a picture of hadrons closely resembling  the structure of physical states in DRM. The hope was strong, but the resulting structure functions were not reasonable for {\it wee} partons \cite{musto}. 

The distance between the DRM, with its fundamental length, and scale invariant deep inelastic physics was not bridged. And the discrepancies between DRM and the actual phenomonology of strong interactions were still present. As a phenomenological model DRM was too rigid. But this tight, beautiful theoretical structure  was so attractive and the emerging string picture was so promising. So,  back in Italy, I joined forces with a group of young physicists, all connected in one way or another with Fubini, all stubbornly convinced that DRM had a rich future.  It was an unusually  large research group, with great enthusiasm and friendship, very much in the equalitarian spirit of the time. I wish only to mention the first paper of the collaboration \cite{Ademollo}, where we extended the string picture of DRM from the spectrum to the interaction, for which only few results had been obtained up to that point. The question was clearly stated: ``{\it If string is not only an anologue model of the DRM one should be able to understand the interaction among the hadrons by starting from the string picture.}". We gave a positive answer to the question,  by showing that the possible interactions were determined by the reparametrization invariance of the string world-sheet.

I like to look at this result as a  step, at least for me, and I think, for all the ones involved in that collaboration, away from an S-matrix approach toward a geometrical space-time description, very much in the spirit of einsteinian physics.
The decisive step in this direction was, obviously, the discovery that the Virasoro-Shapiro dual model contains Einstein's theory of gravity \cite{Yoneya} and Scherk and Schwarz proposal of  a DRM approach to gravity \cite{Scherk}, which opened the way to the development of string theory as a unified theory of all interactions. 
Then one can look to the birth of string theory  as  the growing, inside the shell of the S-matrix approch, of a new geometrical program that is the closest realization of the einsteinian dream.

One may be tempted to synthetize the development of string theory with the motto: ``From Heisenberg to Einstein". But such a simple statement, while  conveying an important part of the story, would require a longer discussion. This not only because Heisenberg had moved away, as we mentioned, from the S-matrix approach to unified field theory, so that he could have liked string, with its fundamental dimensionful constant and excitations corresponding to observable particles. But, because the story is much more complex, let me just mention two points. Notice that,  while the string theory approach gives a simple and intuitive picture, the operatorial approach derived in the DRM framework appears complete and  still today very useful in performing explicit calculations\footnote{On this point see ref. \cite{Schwimmer}.}. And, finally, there is the issue of quantum mechanics. String theory is a full quantum theory and this is very different from Einstein's expectations, even if new developments, such  as classical supergravity-quantum gauge theory duality, could put an old discussion on a new footing.
\vspace{2cm}
\begin{center}
\bf Acknowledgments
\end{center}
I am grateful to the organizers of the Meeting ``The Birth of String Theory'', held in Florence, May 17-18 2007, that allowed me to meet old friends and recall the old days. I wish to thank Dr. Vincenzo De Luise, librarian of my Department, for his constant help.

\end{document}